\documentclass[,final,numberedheadings]
  {aipproc}

\layoutstyle{6x9}

\usepackage{amsmath}
\usepackage{amssymb}
\usepackage{natbib}

\newcommand{\fermi}{\emph{Fermi}}
\newcommand{\DG}{$^{\circ}$}
\newcommand{\al}{$\alpha$}
\newcommand{\ze}{$\zeta$}
\newcommand{\g}{$\gamma$}

\begin{document}

\title{Modeling the Pulse Profiles of Millisecond Pulsars in the Second LAT Catalog of \g-ray Pulsars}

\classification{97.60.Gb}
\keywords      {gamma rays: observations, pulsars: millisecond}

\author{T.~J.~Johnson}{
  address={National Research Council Research Associate, National Academy of Sciences, Washington, DC 20001, resident at the Navel Research Laboratory, Washington, DC 20375, USA, tyrel.j.johnson@gmail.com}
}

\author{A.~K.~Harding}{
  address={Astrophysics Science Division, NASA Goddard Space Flight Center, Greenbelt, MD 20771, USA}
}

\author{C.~Venter}{
  address={Centre for Space Research, North-West University, Potchefstroom Campus, Private Bag X6001, Potchefstroom 2520, South Africa}
}

\author{J.~E.~Grove}{
  address={Space Science Division, Naval Research Laboratory, Washington, DC 20375-5352, USA}
}

\author{the \fermi{} LAT Collaboration and Pulsar Timing Consortium}{address={Across the world.}}


\begin{abstract}
Significant \g-ray pulsations have been detected from $\sim$40 millisecond pulsars (MSPs) using 3 years of sky-survey data from the \fermi{} LAT and radio timing solutions from across the globe. We have fit the radio and \g-ray pulse profiles of these MSPs using geometric versions of slot gap and outer gap \g-ray emission models and radio cone and core models. For MSPs with radio and \g-ray peaks aligned in phase we also explore low-altitude slot gap \g-ray models and caustic radio models. The best-fit parameters provide constraints on the viewing geometries and emission sites. While the exact pulsar magnetospheric geometry is unknown, we can use the increased number of known \g-ray MSPs to look for significant trends in the population which average over these uncertainties.
\end{abstract}

\maketitle

\section{INTRODUCTION}\label{intro}
Millisecond pulsars (MSPs) are thought to be old pulsars which have reached very short spin periods via accretion from a companion \cite{Alpar82}.  Non-recycled pulsars have been known to emit \g-rays for some time but no significant high-energy (HE, $\geq$ 0.1 GeV) pulsations had been seen from any MSP  prior to the launch of the \emph{Fermi Gamma-ray Space Telescope} in 2008.  
Using observations with the Large Area Telescope (LAT, the main instrument aboard \fermi{}, \cite{LAT}) MSPs have been established as a significant class of HE emitters via the detection of pulsed \g-rays, at the radio period, from $\sim$40 MSPs\footnote{For a list of publicly announced detections see:\\https://confluence.slac.stanford.edu/display/GLAMCOG/Public+List+of+LAT-Detected+Gamma-Ray+Pulsars} \cite{2PC}.

The geometry and location of emission in the pulsar magnetosphere remain important and open questions in \g-ray pulsar physics.  Thus, we have generated geometric simulations and developed a likelihood fitting procedure to constrain the viewing geometries and properties of the emission regions of LAT-detected MSPs.  By fitting a large sample of MSPs we can marginalize over discrepancies in the field geometry and still make meaningful statements about the population of \g-ray MSPs as a whole.

\section{LIGHT CURVE SIMULATIONS AND FITTING}\label{simsandfit}
We assume a vacuum retarded-dipole magnetic field geometry \cite{Deutsch55} when simulating pulsar light curves, but note that this geometry is only an approximation as charges will be pulled from the stellar surface and populate the magnetosphere \cite{GJ69}.  Simulated light curves are generated as described in \cite{Venter09,Venter12} using geometric outer gap (OG), slot gap/two-pole caustic (TPC), pair-starved polar cap (PSPC), and low-altitude slot gap (laSG) models with either a single-altitude, hollow-cone and/or core beam; altitude-limited (alTPC/OG), or laSG radio models.

We use a resolution of 1\DG{} in magnetic inclination (\al) and viewing angle (\ze), both with respect to the pulsar spin axis, and of 2.5\% of the polar cap opening angle in emission gap width.  For the alTPC/OG models we use steps of 0.05 light-cylinder radii (R$_{\rm LC}$) in emission altitude.  The laSG models have a resolution of 0.2 stellar radii (R$_{\rm NS}$) in the fading radius and 0.1 (0.3) R$_{\rm NS}$ in the inner (outer) fading parameter (see \cite{Venter12} for definitions).  Accelerating particles are never followed beyond a spherical radius of 1.2 R$_{\rm LC}$ or a cylindrical radius of 0.95 R$_{\rm LC}$.

We fit the simulated \g-ray light curves to the data using Poisson likelihood and the radio using a $\chi^{2}$ statistic and then combine the two.  We scan over the parameter phase space and estimate uncertainties from either 1- or 2-D likelihood profiles.

\section{RESULTS}\label{res}
We have begun fitting the light curves of  MSPs in the second LAT catalog of \g-ray pulsars (2PC, see \cite{2PC} and $\ddot{\rm O}$.~\c{C}elik et al., \emph{these proceedings}).  We select events from the 2PC data set within 2\DG{} of each MSP and having reconstructed energies $\geq$ 0.1 GeV.  The best-fit spectral models are used to estimate the background levels taking into account other sources in the region.

\begin{figure}
\includegraphics[height=.4\textheight]{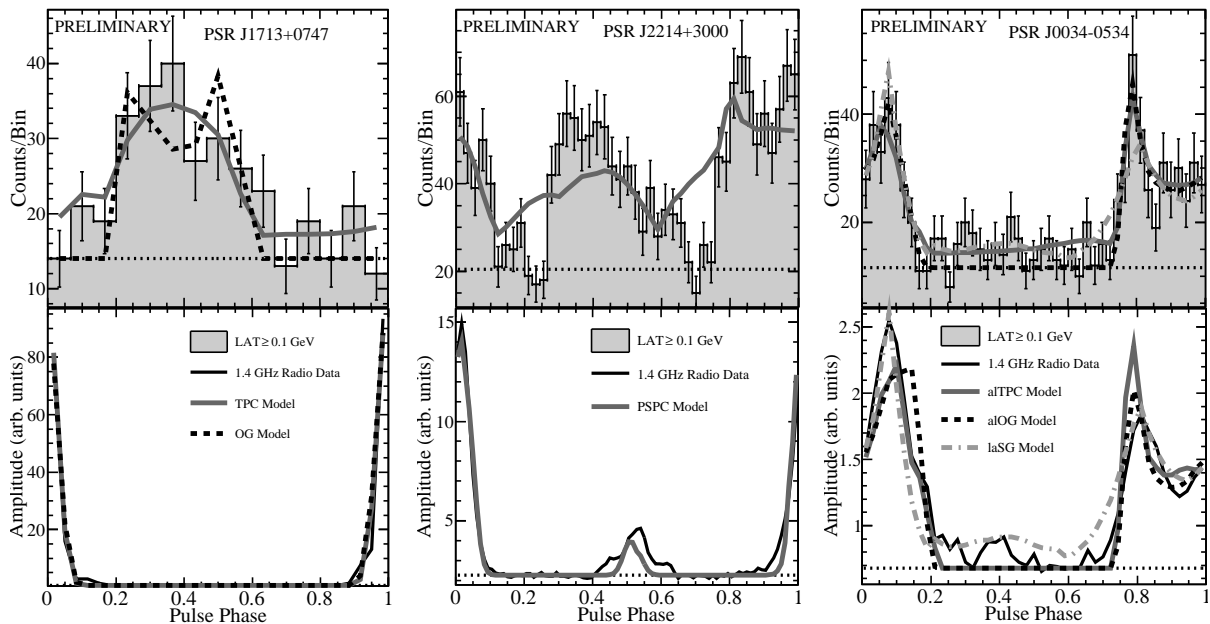}
\caption{Three example light curve fits.  The top panels show the observed and best-fit \g-ray light curves.  The bottom panels show the observed and best-fit radio light curves. Estimated background levels are indicated by dashed, horizontal lines.  The model light curves are indicated in the legend for each MSP.\label{exfits}}
\end{figure}

Fig.~\ref{exfits} shows fits for three \g-ray MSPs, all of which demonstrate the importance of using not just the radio profile but also the polarization information.

The left panel in Fig.~\ref{exfits} shows the observed and best-fit light curves of PSR J1713+0747 using the TPC and OG \g-ray models with a hollow-cone plus core radio beam.  Polarimetric observations detect sense-changing circular polarization through the main pulse of this MSP \cite{Stairs99} which suggests a core component.  Neglecting the core component in the modeling can lead to differences in \al{} and \ze{} of as much as 30\DG{}.

The center panel in Fig.~\ref{exfits} shows the same for PSR J2214+3000 in which the \g-rays are fit with a PSPC model and the radio with a hollow-cone beam.  This MSP had been fit with alTPC/OG models previously \cite{thesis}; however, recent observations detect significant linear polarization (not expected for caustic emission \cite{Venter12}) and with more statistics the \g-ray peaks clearly lead those in the radio, all arguing against the alTPC/OG models.

The right panel in Fig.~\ref{exfits} shows the same for PSR J0034$-$0534, an MSP with \g-ray and radio peaks aligned in phase.  Observations of this MSP are consistent with 0\% linear polarization \cite{Stairs99} which argue for caustic radio models (i.e., alTPC/OG).  Note that the polarization properties of the laSG model have not been investigated.

\begin{figure}
\includegraphics[height=.5\textwidth]{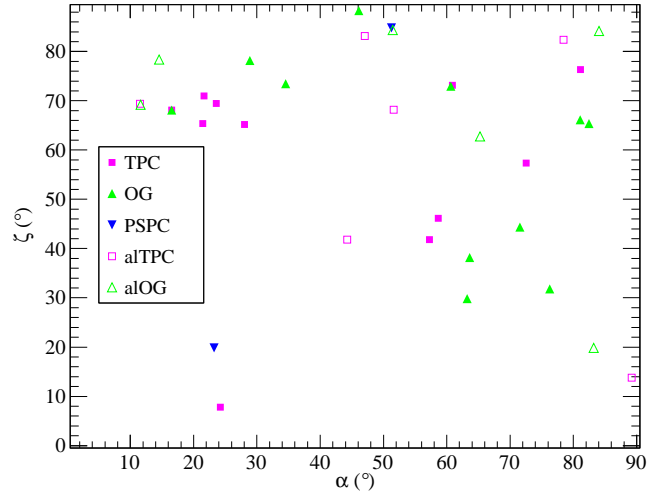}
\caption{Best-fit (\al,\ze) pairs for each MSP in \cite{thesis}, corresponding emission models are indicated by the legend.  Note that there are two entries for those MSPs fit with TPC and OG variants.\label{az}}
\end{figure}

A similar study of MSP light curves has been performed using two years of LAT data and only 19 MSPs \cite{thesis}.  The best-fit \al{} and \ze{} values from this analysis are shown in Fig.~\ref{az}.

While the \ze{} values are consistent with a random distribution of spin-axes with respect to our line of sight (preference for values closer to 90\DG{} where there is more solid angle) the best-fit \al{} values seem to favor all angles equally.  This is in conflict with studies which suggest that the spin and magnetic axes should align as a pulsar spins down \cite{Young10} or during the recycling process \cite{Ruderman91} and may indicate that the magnetic axis in MSPs (which have weaker magnetic fields) moves towards anti-alignment while spinning down, after the recycling process has completed.

\section{FUTURE}\label{fut}
Once we have fit the profiles of all MSPs in the 2PC we will look for population trends in viewing geometry, energetics, and best-fit model with a larger sample than what has been done previously \cite{thesis}.  We will also generate similar simulations using different magnetospheric geometries, explore radio cone/core models at altitudes different than those in \cite{Story07}, and investigate mixtures of caustic and non-caustic radio emission.

\begin{theacknowledgments}
The \fermi{} LAT Collaboration acknowledges support from a number of agencies and institutes for both development and the operation of the LAT as well as scientific data analysis. These include NASA and DOE in the United States, CEA/Irfu and IN2P3/CNRS in France, ASI and INFN in Italy, MEXT, KEK, and JAXA in Japan, and the K.~A.~Wallenberg Foundation, the Swedish Research Council and the National Space Board in Sweden. Additional support from INAF in Italy and CNES in France for science analysis during the operations phase is also gratefully acknowledged.  Portions of this research performed at NRL are sponsored by NASA DPR S-15633-Y.
\end{theacknowledgments}

\end{document}